\documentclass[12pt,a4paper]{article}
\usepackage[latin1]{inputenc}
\usepackage{amsmath}
\usepackage{amsfonts}
\usepackage{amssymb}
\usepackage{graphicx}
\usepackage{calc}
\usepackage[a4paper,margin=1.0in]{geometry}
\usepackage{subfig}

\usepackage{cite}
\usepackage{hyperref}
\newcommand{\ee}{\ensuremath{\mathrm{e}^+\mathrm{e}^-\,}}
\newcommand{\ie}{{\it{i.e.}}}
\newcommand{\eg}{{\it{e.g.}}}

\begin{document}
\title{Updated status of the undulator-based ILC positron source}
\author{
S. Riemann$^1$\thanks{corresponding author, sabine.riemann@desy.de}, P. Sievers$^2$,  
G. Moortgat-Pick$^3$, A. Ushakov$^3$\\
\\ $~$ 
 \normalsize $^1$\textit{Deutsches Elektronen-Synchrotron (DESY), Platanenallee 6, D-15738 Zeuthen}\\
 \normalsize $^2$\textit{CERN, CH-1211 Geneva 23, Switzerland}\\
 \normalsize $^3$\textit{University of Hamburg, Luruper Chaussee 149, D-22761 Hamburg}}
 \date{   }
\maketitle
\begin{abstract}
\noindent
The  design of the positron source for the International Linear Collider (ILC) is still under discussion. The baseline design plans to  use the high-energy electron beam for the positron production before it goes to the IP. The  electrons pass a long helical undulator and  generate an intense circularly polarized photon beam which hits a thin conversion target to produce \ee{} pairs.  The resulting positron beam is longitudinally polarized which provides an important benefit for  precision physics analyses at the ILC.  
In this paper the status of the positron target design  studies is presented. Focus is the positron yield for center-of-mass energies of 250\,GeV and also the Z peak.  Possibilities to improve the positron collection system and thus to increase the positron yield  are discussed.  
\end{abstract}.
\section{Introduction}\label{sec:intro}

The discussion on the  positron production scheme for a high-energy linear \ee collider is ongoing: Two schemes are under consideration:  The baseline scheme using a helical undulator passed by the high-energy electron beam to generate an intense photon beam for the positron production in a thin target, and  
the scheme based on the use of a separate electron beam to create \ee pairs in a thick target.  
The efficiency of positron production in a conversion target together with the capture acceleration of the positrons is low, so in both cases it is a challenge to generate  the  $1.3\times 10^{14}$ positrons per second  that are required at the ILC collision point  (nominal luminosity).  However, using a helical undulator allows to produce a circularly polarized photon beam for the generation of a longitudinally polarized positron beam.  A high-energy \ee{} collider with both beams polarized broadens substantially the physics potential and has been chosen as baseline option for the ILC positron source~\cite{TDR1,TDR31,TDR32}.

In this paper the status of the studies for the undulator based source is updated; it continues the results given in references~\cite{ref:LCWS18} and~\cite{ref:e+WG}. Details on the electron driven source can be found in references~\cite{Omori:2011wq,Nagoshi:2020blm}.

This paper is organized as follows: The basic  parameters of the undulator source are given in section~\ref{sec:sourcepar}. 
Due to the low  efficiency of the \ee{} pair production  and the required high luminosity, the heat load in the  target and hence the cooling are important issues. To stand the heat load the positron target must be moved and is designed as rotating wheel. The basic items are discussed in detail in references~\cite{ref:LCWS18} and~\cite{ref:e+WG} and summarized in section~\ref{sec:sourcepar}.   
The positron yield depends strongly on the design of the optical matching  device (OMD) located directly behind the conversion target. In section~\ref{sec:OMD}   the influence of the  magnetic field on the positron  yield is discussed. It is well known, that a high magnetic field at the target increases substantially the positron yield. Possible options to realize the magnetic field required for high positron yield as well as the resulting eddy currents in the spinning target which could increase the heat load are  reviewed in section~\ref{sec:OMDdesign}.   The necessary R\&D work is shortly outlined in section~\ref{sec:RandD}. 
Section~\ref{sec:e+pol} emphasizes the  benefit of a polarized positron beam for the ILC project.  A short summary is given in section~\ref{sec:sum}.   
\section{The undulator-based  ILC positron source}\label{sec:sourcepar}
\begin{figure}[htbp]
\center
 \includegraphics*[width=120mm]{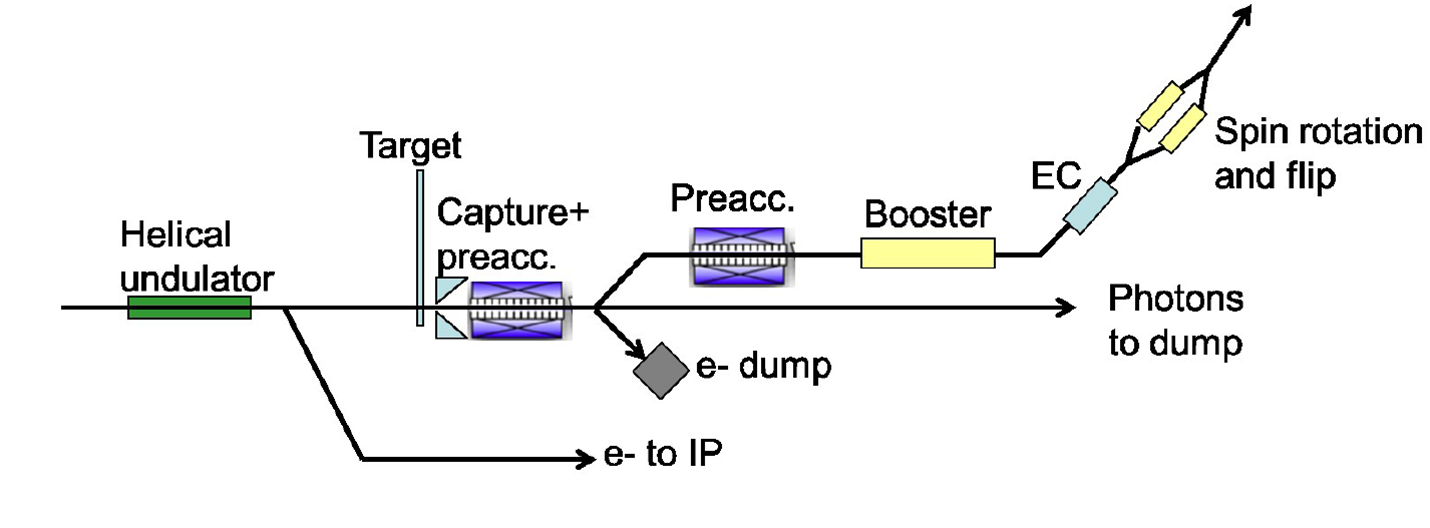}
  \caption{Sketch of the undulator-based ILC positron source.} 
 \label{fig:source}
\end{figure}

The ILC positron source is located at the end of the main linac. It consists of the helical undulator with maximum active length of 231\,m to generate the photon beam, the thin conversion target made of Ti6Al4V, the optical matching device and the capture optics, acceleration, energy and bunch compression, spin rotation and spin flipper as shown in figure~\ref{fig:source} and references~\cite{TDR1,TDR31,TDR32,Aihara:2019gcq}. 
Goal is a positron yield of 1.5 e$^+$/e$^-$ at the damping ring.
Since the photon energy and yield, and hence the positron yield depend strongly on the electron energy, the source performance  has to be studied and optimized for each centre-of-mass energy. 

\subsection{The helical undulator}\label{sec:helund}
A superconducting helical undulator is used to generate the circularly polarized photon beam. A  prototype was manufactured and tested in UK~\cite{ref:undproto}. It consists of two 1.75\,m long undulator modules inserted in a 4\,m long cryomodule. The undulator period is $\lambda_{\mathrm{u}}=11.5\,$mm and the maximum B field is 0.86\,T corresponding to a  value $K=0.92$. As described in the TDR, with 132 undulator modules (66 cryomodules) an active undulator length of 231\,m is reached. Every 3 cryomodules    quadrupoles are foreseen so that  the undulator system reaches a total length of 320\,m. 
The given undulator period, $\lambda_{\mathrm{u}}$, and the maximum field on axis, $B_0$, define together with te electron beam energy the possible parameter range  for the positron source:
The undulator $K$ value is $K\propto \lambda_{\mathrm{u}} B_0$.
The efficiency of positron generation in the target depends on the pair production cross section and hence on the photon energy. The cut-off for the first harmonic is related to the electron energy $E_\mathrm{e}$, $K$ and $\lambda_{\mathrm{u}}$ by
\begin{equation}
E_{1 \gamma} \propto  \frac{E_{\mathrm{e}}}{\lambda_{\mathrm{u}}(1+K^2)}\,,\label{eq:Egamma}
\end{equation}
i.e. lower K values increase the photon energy.
The number of photons created per undulator length is 
\begin{equation}
N_{\gamma} \propto  \frac{K^2}{\lambda_{\mathrm{u}}(1+K^2)}\,,\label{eq:Ngamma}
\end{equation}
implying that low $K$ values result in less photons.   
For an electron beam energy 125\,GeV   a high $K$ value and the full active length of the undulator are necessary  to get the required number of positrons. 
\\
Also the beam spot size on the target depends on the opening angle of the photon beam,
\begin{equation}
\theta_{\gamma} \propto  \frac{\sqrt{1+K^2}}{\gamma}\,,\label{eq:theta}
\end{equation}  
and it is very small even at a large distance from the undulator. This defines the design for the conversion target;  during the 0.73\,ms (baseline) up to 1\,ms long bunchtrain   overheating must be prevented to avoid serious damage. 
Details are discussed in section~\ref{sec:target}.

\subsection{The conversion target}\label{sec:target}
The narrow photon beam causes a high peak energy deposition density (PEDD) in the target material. To prevent overheating during one ILC pulse, the target is designed as wheel of 1\,m diameter spinning with 100\,m/s circumferential speed to distribute the beam load over about 7-10\,cm. The target thickness should be optimized regardig the energy of the electron beam; for ILC250 a thickness of 7\,mm Ti6Al4V is recommended ($\approx 0.2\,$X$_0$). Downstream the target, an optical matching device (OMD) collects the  positrons  using a high B field which goes adiabatically down to 0.5\,T at the accelerating structures. To achieve the required positron yield and to protect the accelerating structures the target wheel rotates in vacuum. 
The PEDD and the average power deposited in the target as well as the positron yield  vary for different $E_\mathrm{cm}$; they also depend on the distance  between target and undulator. 
In previous studies the interplay of the source parameters has been studied for different centre-of-mass                                 energies~\cite{ref:LCWS17proc,ref:LCWS18,ref:e+WG,ref:posipol16-AU,ref:AU-250GeVthickness,ref:AU-OMDyield}.
Most of these studies assumed a pulsed flux concentrator (FC) as optical matching device (OMD). A promising prototype study for the FC was performed by LLNL~\cite{ref:Gronberg-FC}. However, detailed studies identified some weak items of this design: The B field distribution along z cannot be kept stable over the long bunch train duration. So the luminosity would vary during the pulse which is not desired. Further, the particle shower downstream the target causes a high load at the inner part of the flux concentrator front side, which is at least  for ILC250 beyond the recommended material load level~\cite{ref:AU-OMDyield}. This is mainly caused by the larger opening angle of the photon beam   
and the wider distribution of the shower particles downstream the target  
at ILC250. 
To resolve this problem, the drift space between the middle of undulator and the target was reduced to 401\,m. In addition,  
a quarter wave transformer is suggested as OMD since it has a larger aperture.     Further details can be found in references~\cite{ref:AU-OMD, ref:AU-OMDyield}. An alternative could be a pused solenoid; this option is disussed in section~\ref{sec:OMD}. 

Another important issue is the cooling of the target spinning in vacuum. The previously contemplated water cooling was given up and replaced by cooling by thermal radiation. 

Table~\ref{tab:sourcepar} presents an overview of the relevant parameters for the studies with focus on an 125\,GeV  electron beam for positron production.   
\begin{table}[h]
\begin{center}
\renewcommand{\baselinestretch}{1.2}
\begin{tabular}{|lc|cc|}
\hline
                        &     &   FC & QWT \\ \hline
electron beam energy    & GeV &  \multicolumn{2}{c|}{126.5}\\
undulator active length &  m  & \multicolumn{2}{c|}{231} \\ 
space from middle of undulator to target & m & \multicolumn{2}{c|}{401} \\ \hline
undulator K             &     & 0.85& 0.92 \\
photon yield per m undulator & $\gamma$/(e$^- \, $m) & 1.70 &  1.95 \\
photon yield            & $\gamma$/e$^-$     &392.7&  450.4\\ 
photon energy (1$^\mathrm{st}$ harmonic) & MeV & 7.7  & 7.2\\ 
average photon energy                  & MeV & 7.5  & 7.6 \\
average photon beam power              &  kW & 62.6 & 72.2\\
average power deposited in target       & kW &  1.94 & 2.2 \\
rms photon beam spot size on target ($\sigma$) & mm &  1.2 &1.45 \\
PEDD in target per pulse (100\,m/s)       & J/g & 61.0 & 59.8\\ 
\hline 
\end{tabular}
\caption{\label{tab:sourcepar}Summary of the source performance parameters for ILC250 with  1312 bunches per pulse and two different undulator K values. The pulse repetition rate is 5Hz. The numbers are shown for a decelerating capture field.  See also references~\cite{TDR1,TDR31,TDR32,Aihara:2019gcq,ref:posipol16-AU,ref:AU-250GeVthickness,ref:AU-OMDyield}. }
\end{center}
\end{table}

The positron yield  was simulated for a flux concentrator  and a quarter wave transformer (QWT). In both cases the positron beam  polarization is 30\%.
However, the positron yield depends strongly on the magnetic field assumed for the simulations. 
The numbers in table~\ref{tab:sourcepar} given for the QWT suppose an optimized shape of the B field: At the target exit the B field is zero. It achieves the maximum value of 1.04\,T at a distance of 7.6\,mm downstream the target. However, the  design  given in reference~\cite{ref:GaiLiu-OMD} corresponds to a realistic design and has the maximum B field at a distance of about 3.5\,cm downstream the target exit. So for the ideal case the design yield of $\approx 1.5$e$^+$/e$^-$ is almost achieved while with the realistic QWT the yield is less than but close to 1. More details are given in section~\ref{sec:OMD}.
\subsubsection{Temperature distribution in the target wheel}\label{sec:wheel}
\begin{figure}[htbp]
\center
 \includegraphics*[width=130mm]{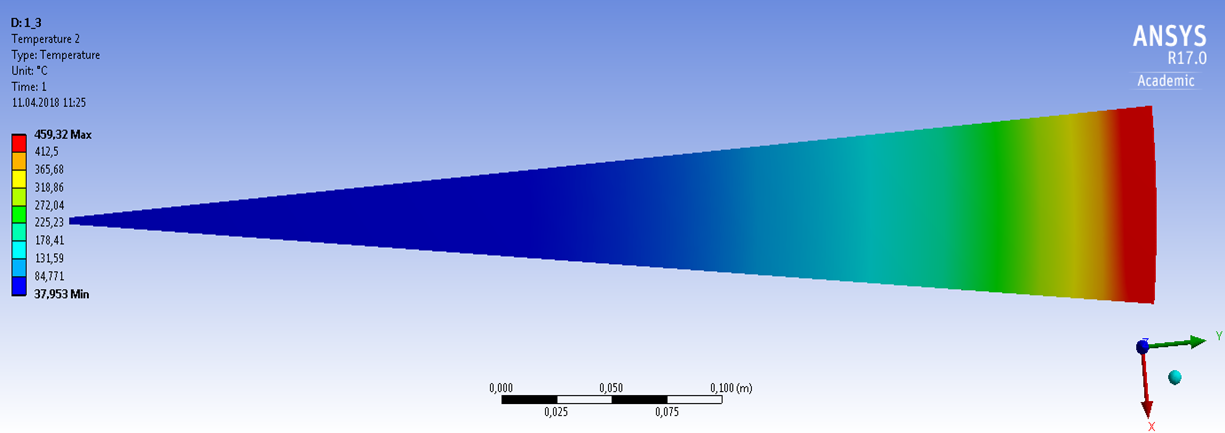}
  \caption{ Average temperature distribution in  the target shown for a sector corresponding to 1 pulse length (0.73\,ms) at ILC250; the beam impinges on the target at r=50\,cm. The emissivities of target and cooler surface are 0.5 corresponding to an effective emissivity of $\varepsilon  = 0.33$.}   
 \label{fig:Tsector}
\end{figure}

The average energy deposition in the ILC positron target is about $2-7\,$kW depending on the drive beam energy in the undulator, the target thickness and the luminosity (nominal or high). 
For ILC250, the average energy deposition in the target is 2\,kW. 
\\
Since the initial investigations of the wheel, involving leak tight rotating vacuum seals and water cooling showed major problems~\cite{ref:Gronberg-FC,ref:Gronberg-posipol13}, an alternative technical solution was brought up to ensure the heat radiation as well as the safe rotation of 2000\,rpm by magnetic bearings~\cite{ref:sievers-posipol14,ref:sievers-posipol16}. This proposal is now considered as solution for the target wheel design. 
The energy deposition of few kW can be extracted by radiation cooling if the radiating surface is large enough and the heat distributes fast enough from the area of incident  beam to a large radiating surface.  
The wheel spinning in vacuum radiates the heat to a stationary  cooler opposite to the wheel surface. It is easy to keep the stationary cooler at room temperature by water cooling.
But it is crucial for the design that the  heat distributes from the volume heated by the photon beam  to a larger surface area. 
The thermal conductivity of Ti6Al4V is low, $\lambda = 0.068\,$W/(cm\,K) at room temperatures and  0.126\,W/(cm\,K) at 540$^\circ$C. The heat capacity is $c=0.58\,$J/(g\,K) at room temperature and 0.126J/(g\,K) at 540$^\circ$C~\cite{Mills:thermo}. With the wheel rotation frequency of 2000rpm   each part of the target rim is hit after 6-8 seconds but this time is not sufficient to distribute the heat load almost uniformly over a large area.  
The heat is accumulated in the rim and the  highest temperatures are located in a relatively small region around the beam path. The average temperature distribution was calculated using the ANSYS software package~\cite{ref:ANSYS} and  is shown in figure~\ref{fig:Tsector} for one sector representative for the track of one bunch train.

\begin{figure}[htbp]
\center
 \includegraphics*[width=150mm]{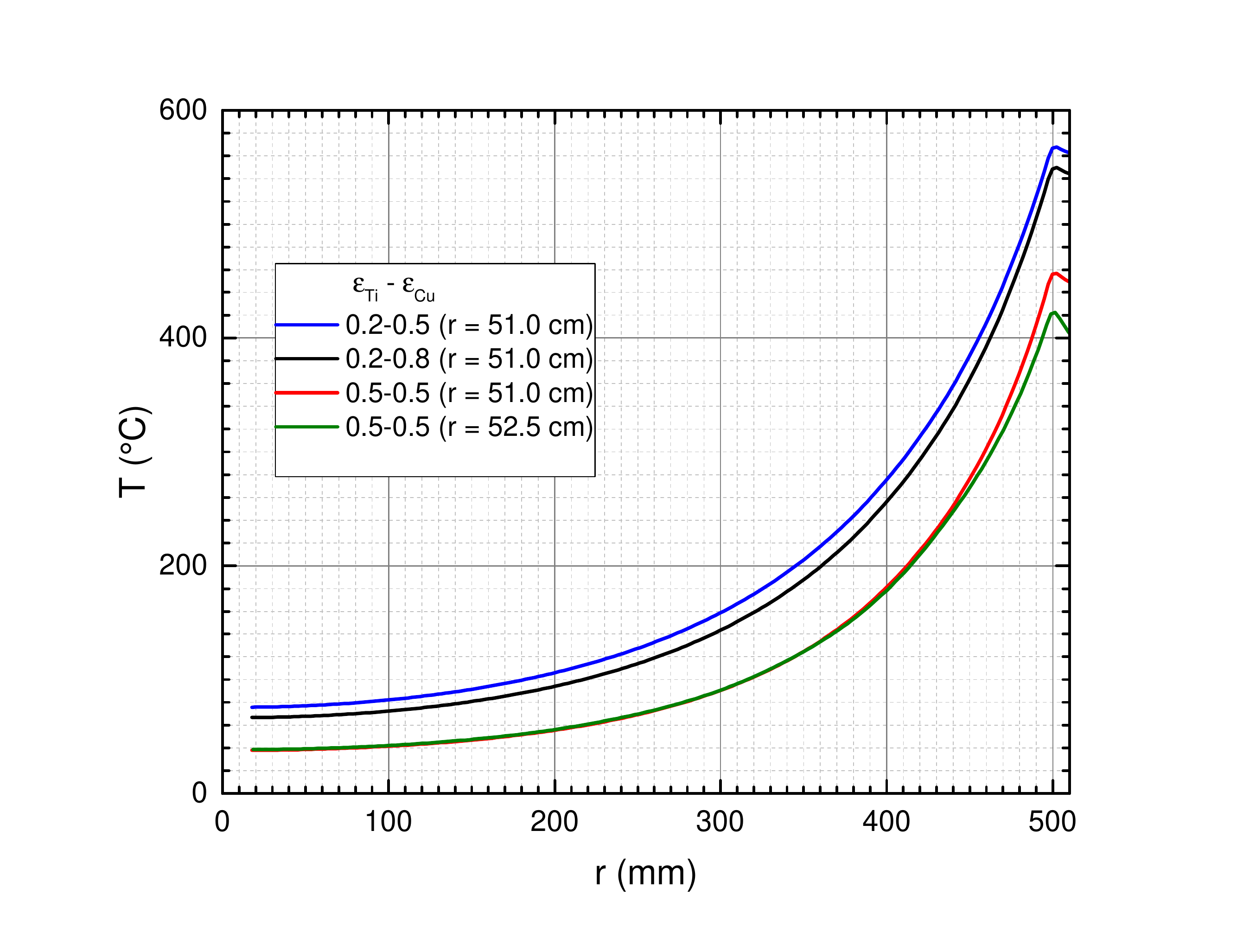}
  \caption{Radial average  temperature distribution in the target wheel for various emissivities of target and cooler. The beam hits the target at  a radius of 50\,cm. The outer wheel radius is 51\,cm and also 52.5\,cm.}
 \label{fig:Trad}
\end{figure}

The temperature depends substantially on the emissivity $\varepsilon$,  of the surfaces.  An optimization of the emissivities by surface processing or coating is possible and should be tested also under long-term irradiation conditions as expected at the ILC. 
Figure~\ref{fig:Trad} shows the radial temperature profile assuming different emissivities. In addition also the radius of the wheel increased to enhance the surface for the   T$^4$ radiation. A rough estimate~\cite{ref:LCWS18,ref:e+WG} shows that with a wheel radius of about 60\,cm the average temperature could be decreased by 100\,K or even more. 
This is important since every 6-7 seconds the beam pulse increases the temperature by approximately 60-100\,K within about 50\,$\mu$s depending on the required luminosity. The material must stand the   cyclic load at elevated temperatures. %
Experimental tests were performed with the electron beam of the microtron in Mainz (MAMI) to simulate the cyclic load  as expected during ILC operation~\cite{Ushakov:2017dha,Heil:2017ump} and demonstrated that the material is suited for this application. 

The thermomechanical stress in a rotating target wheel disc is discussed in detail in references~\cite{ref:LCWS18,ref:LCWS17proc,ref:e+WG}. The results of the irradiation tests at MAMI and detailed simulation studies with ANSYS showed that the expected load at the ILC positron target is below the material  limits~\cite{ref:AU-posipol18,ref:LCWS18}. However, for the final construction of the target wheel FEM design  studies and  systematic tests using a mock-up are necessary  to ensure a long-term operation without failure.

\subsection{Engineering and Design Options for the Rotating Wheel}\label{sec:engin}
For the studies of the positron yield optimization, the temperature distribution and  cooling principle  a target wheel designed as full 1\,m-diameter disc of 7mm thickness  made of Ti6Al4V was assumed. 
As illustrated in figure~\ref{fig:Trad}, the radial steady state temperature in the wheel depends also on the radius. Due to the the heat conductivity in the target material and the $T^4$ dependence of the evacuation by thermal radiation of the power, most of it is evacuated close to the rim of the wheel. For example,  
most power is radiated from the rim at radii larger 35\,cm.
 
\begin{figure}[hp]
\center
 \includegraphics*[width=130mm]{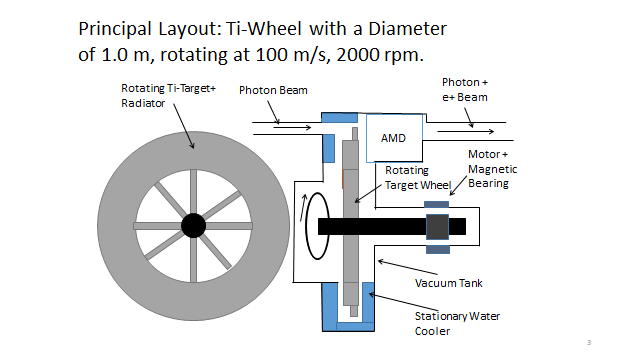}
  \caption{Principal Layout of the rotating wheel showing its main components: its cooling system, its rotating magnetic bearing and the  matching device (AMD). } 
 \label{fig:peter-layout}
\end{figure}

\begin{figure}[hp]
\center
 \includegraphics*[width=130mm]{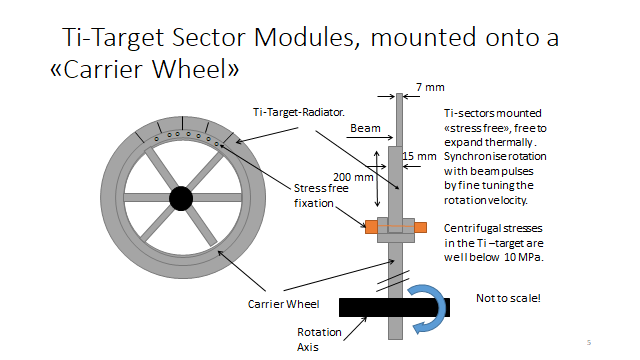}
  \caption{Details of the mechanical layout of the target units, fitted to the carrier wheel. The  extension of the Ti-sheet into both radial directions beyond the beam impact and the increase of its thickness will lead to reduced average temperatures in the target.  } 
 \label{fig:peter-layout-detail}
\end{figure}

By increasing the outer radius of the wheel up to 60 cm, while maintaining the beam impact at r=50 cm, substantially  lower average temperatures can be expected.  
Thus it is possible to  conceive a target wheel consisting of two distinct parts with separate functionalities: A  'carrier wheel', designed and optimized in terms of weight, material,  moment of inertia,  centrifugal forces, stresses and vibrations, etc., and a second unit, the actual Ti-target rim. The target units are fitted mechanically to the rim of the carrier wheel in such a way that the cyclic loads, temperature rises and stresses in the target units are not or little transmitted to the carrier wheel.  This would allow to design and optimize the engineering of the carrier wheel independently from that of the target proper.  A possible layout   in 
Figure~\ref{fig:peter-layout} shows the main items of the target wheel, the spoked rotating carrier wheel with its magnetic bearings and the water cooled stationary coolers. 
In   figure~\ref{fig:peter-layout-detail} some details are suggested to fit the target to the carrier wheel. 

\section{Positron yield and OMD}\label{sec:OMD}
\begin{figure}[h]
\centering
\begin{tabular}{lr}
 \includegraphics*[width=75mm]{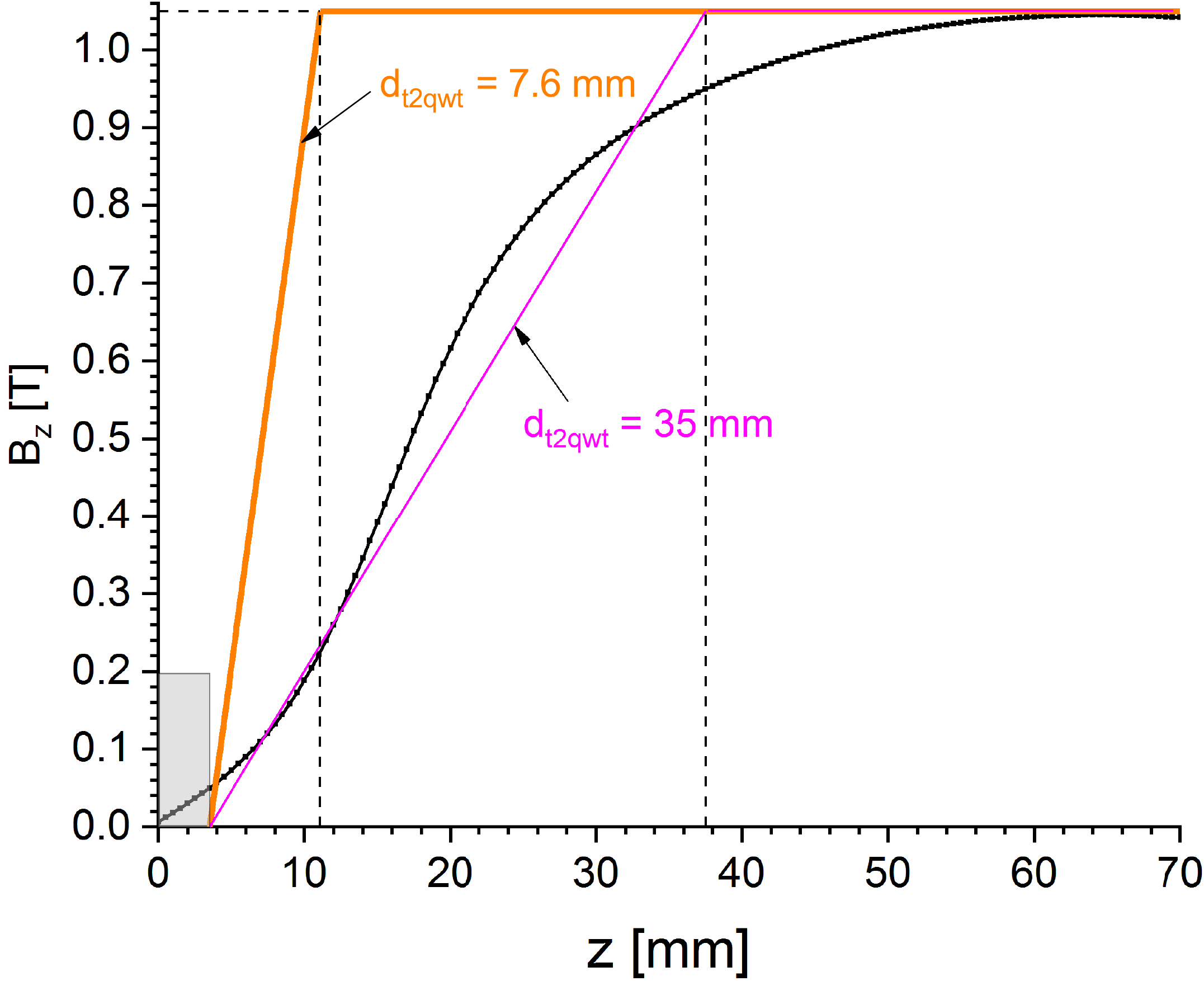}&
 \includegraphics*[width=75mm]{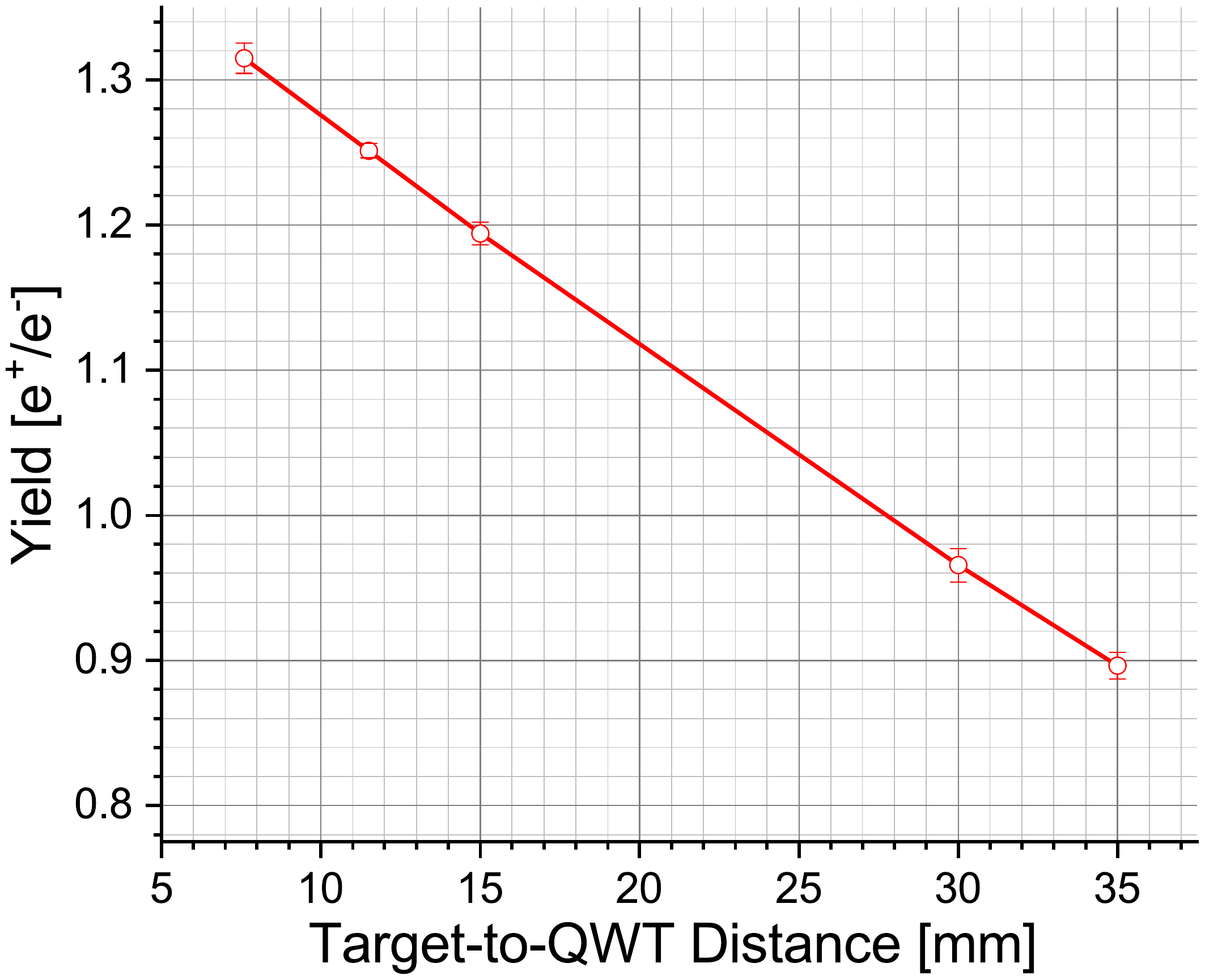}
\end{tabular}
  \caption{{\it{Left:}} Shape of the magnetic field on axis of the QWT. The yellow line shows the field as used for the simulation studies with an 'ideal' OMD to  achieve the required positron yield. The black line gives the B field as suggested for the QWT in reference~\cite{ref:GaiLiu-OMD}. The magenta line is the corresponding approximation used for the simulation of the positron yield. 
{\it{Right:}} Positron yield depending on the distance d$_\mathrm{t2qwt}$ assuming a field distribution as given with the yellow and magenta line of the left plot.  The undulator $K$ value is $K=0.85$.}
 \label{fig:QWT-Bz-Y}
\end{figure}

During the long time of design studies for the ILC positron source several options for the OMD have been considered. An overview is given in reference~\cite{ref:GaiLiu-OMD}.  
The best solution would be a pulsed flux concentrator which initially  was envisaged for the adiabatic matching device (AMD). 
But in further detailed studies it appeared that over the long pulses of about 1\,ms the field in the FC could not be kept stable in time~\cite{ref:e+WG}.
 Further, studies~\cite{ref:AU-OMD} showed that the peak energy deposition in the center at the front of the FC is too high for ILC250. Neither a larger FC aperture nor a shorter distance of FC to  undulator could improve the situation substantially. 
 Therefore, further studies were pursued with a QWT. 
Following reference~\cite{ref:GaiLiu-OMD}, only a yield  $Y\le 1\,$e$^+/$e$^-$ can be reached for electron beam energies of 125\,GeV. Studies~\cite{ref:AU-OMD} showed that for the thinner conversion target, a maximum K value ($K=0.92$) and an optimized B field the yield can be increased to the required value. Figure~\ref{fig:QWT-Bz-Y} demonstrates the influence of the magnetic field shape on the positron yield. If the magnetic field rises from almost zero at the target exit within 8\,mm to a maximum value of 1.04\,T, a yield of about 1.5e$^+$/e$^-$ can be reached by increasing  $K$ to 0.92. 
It is the question whether a QWT with the corresponding optimum magnetic field is possible, \ie{} maximum field of $\approx 1\,$T at a distance of 7--8\,mm from the target and almost zero at the target. 
Studies  have shown that a high magnetic field at the target increases substantially the positron yield. However,  the eddy currents created in  the spinning target are considered as drawback  since they heat the target~\. Even if the heating is acceptable, brake effects 
have to be taken into account.   

\subsection{Magnetic field in target}\label{sec:Yield-Bfield}
The positron yield at low centre-of-mass energies (ILC250, GigaZ) cannot be increased with a longer undulator but with an  optimization of the B field in the matching section  the required yield of 1.5\,e$^+$/\,e$^-$ could be reached.

\begin{figure}[h]
\centering 
\includegraphics*[width=100mm]{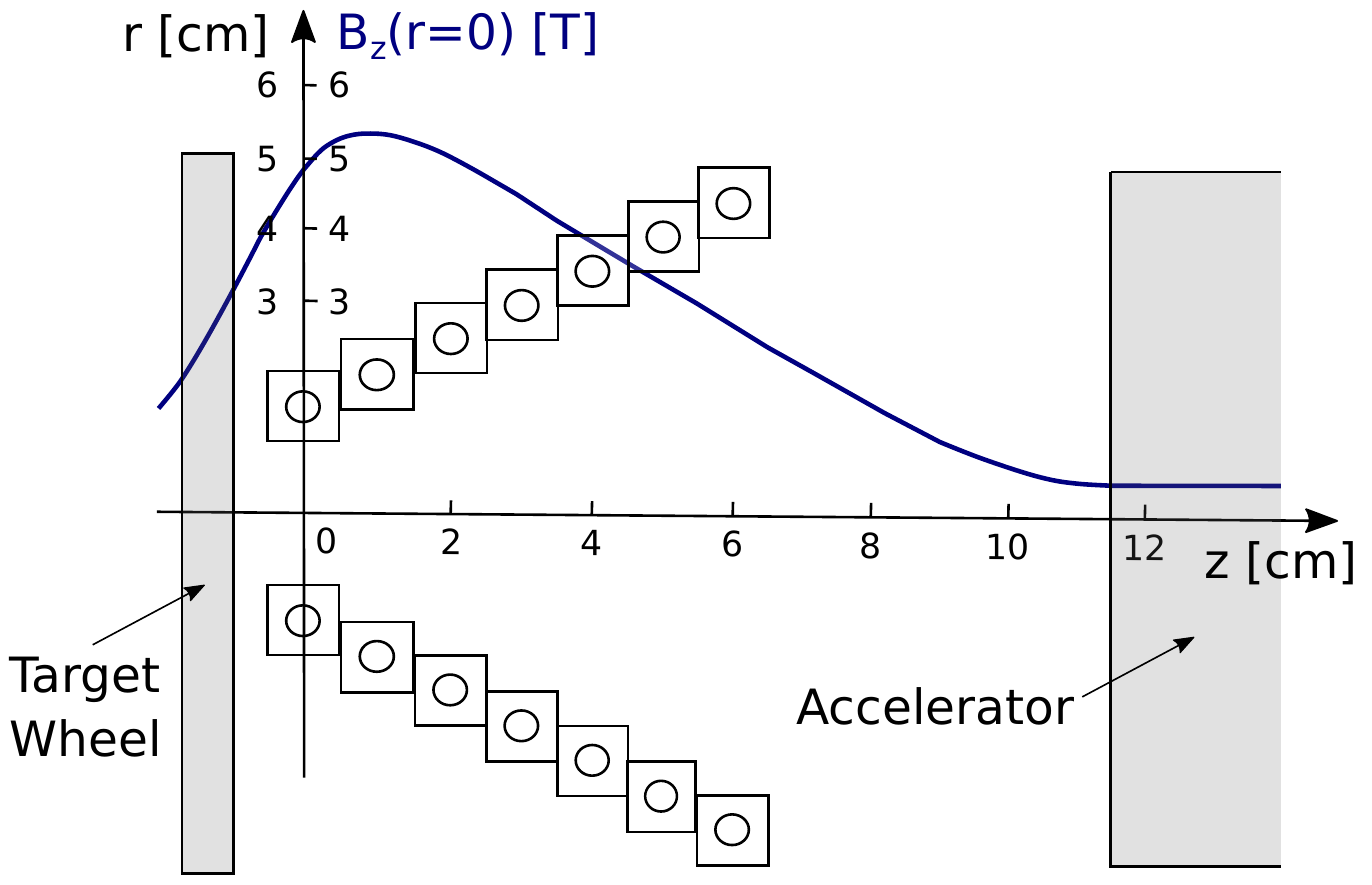} 
\caption{B field as used for the studies to increase the positron yield.}
\label{fig:B+solenoid}
\end{figure}

\begin{figure}[h]
\begin{tabular}{lr}
 \includegraphics*[width=70mm]{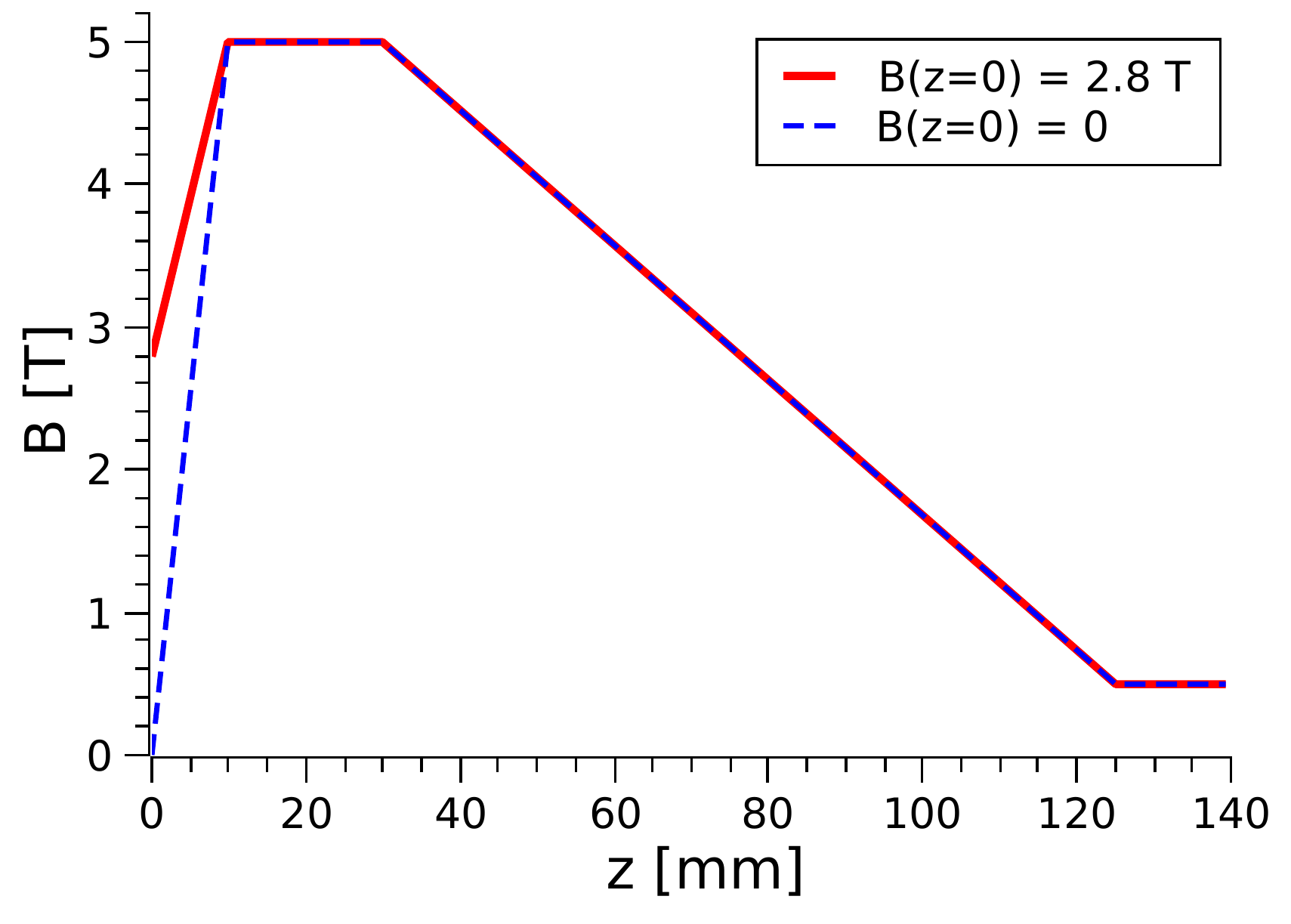}& 
\includegraphics*[width=80mm]{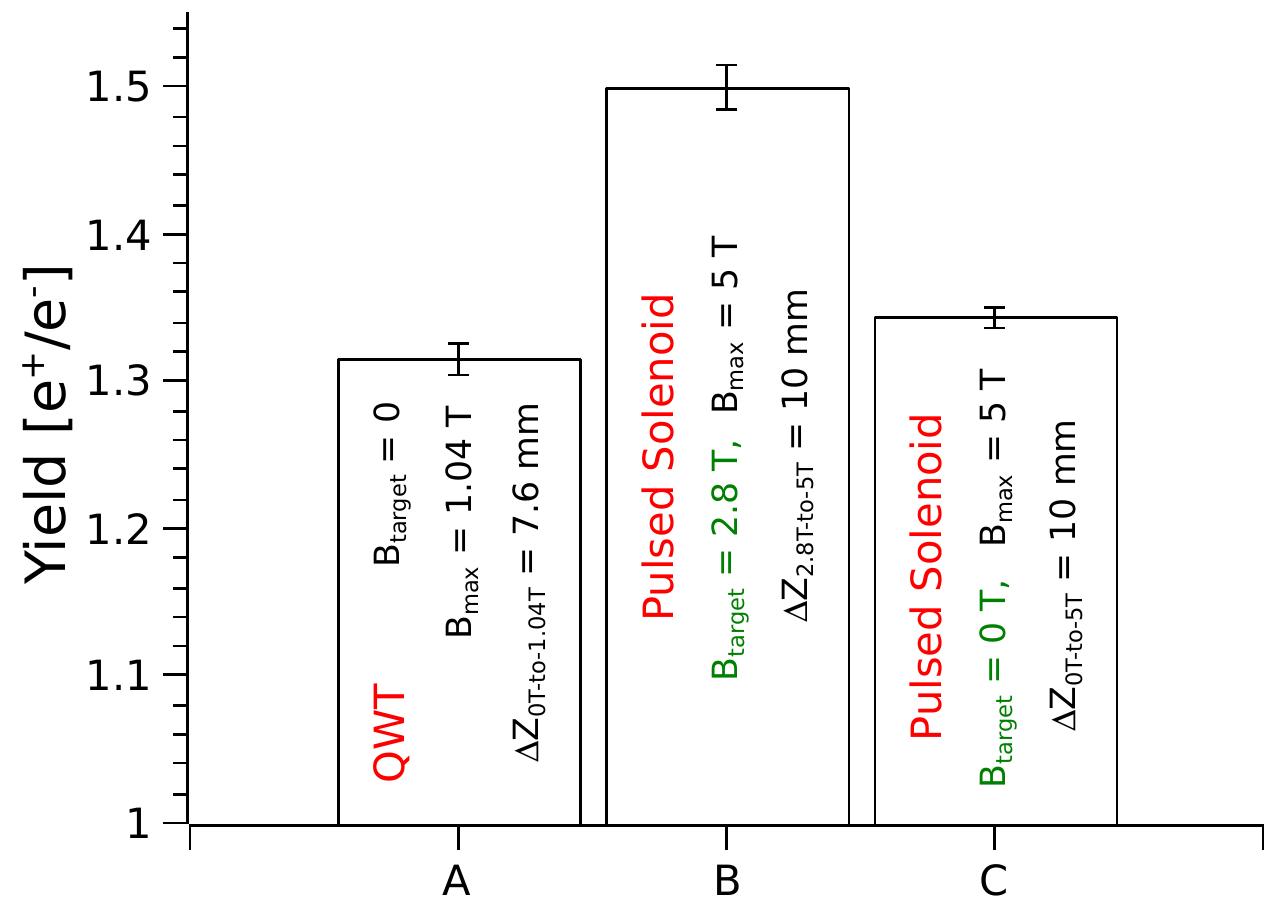}
\end{tabular}
\caption{{\it{Left:}} B field options as used for the studies to increase the positron yield. {\it{Right:}} Comparison of the positron yield for three B field options. }
\label{fig:B+Y-study}
\end{figure}

As a first approach to study the influence of the B field at the exit side of the target on the positron yield the simplified field profiles shown in figure~\ref{fig:B+Y-study}  have been used. Such fields can be achieved in a conical solenoid as sketched in   
figure~\ref{fig:B+solenoid}.

Figure~\ref{fig:B+Y-study} compares the positron yield results assuming three different B field options and demonstrates that the required positron yield can be realized.    

\begin{figure}[h]
\centering 
 \includegraphics*[width=100mm]{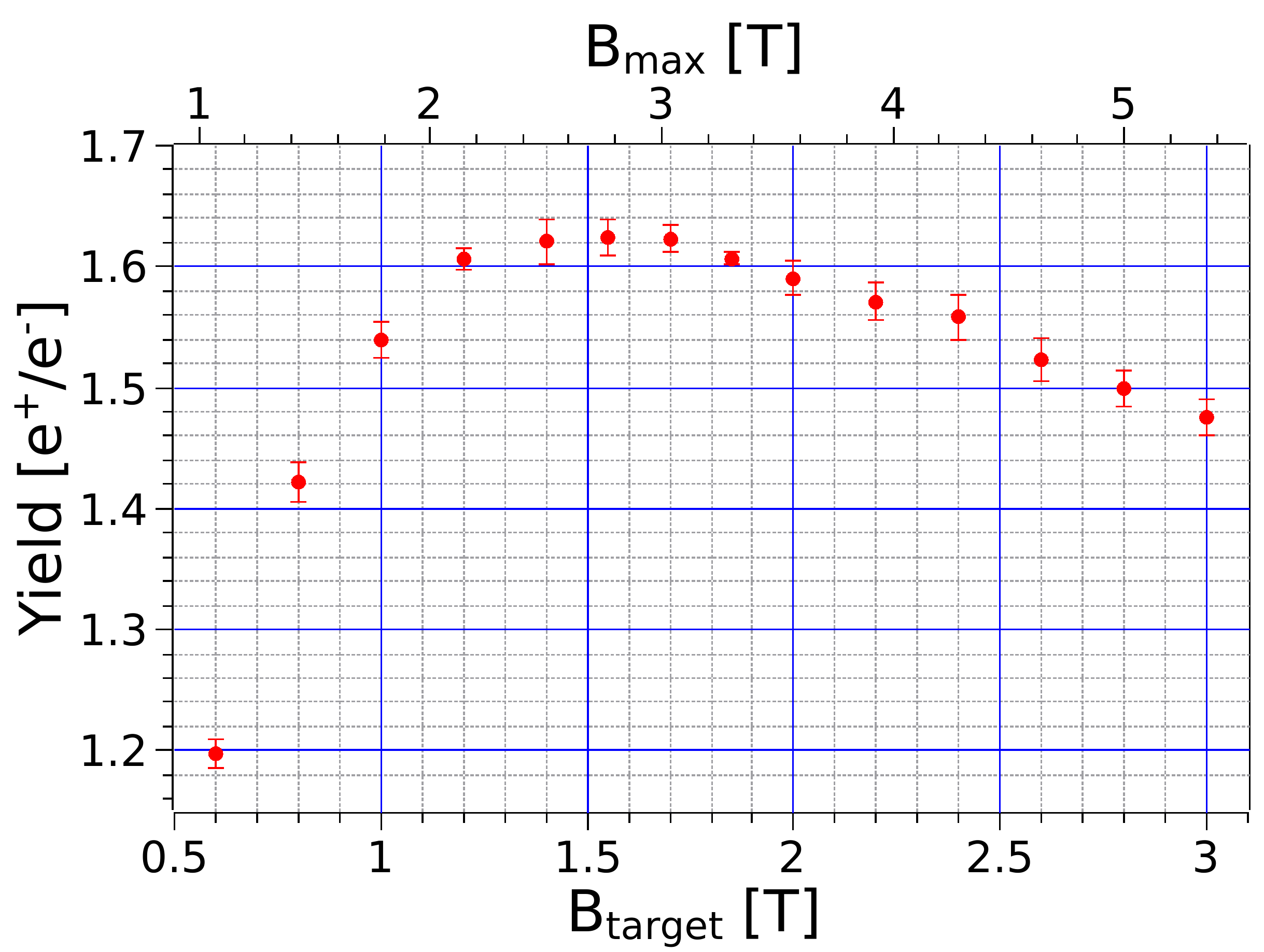}
\caption{Expected positron yield depending on the  field $B_\mathrm{target}$ at the target exit and on the  maximum   field $B_\mathrm{max}$. }
\label{fig:Y-B}
\end{figure}

As demonstrated in figure~\ref{fig:Y-B}, further studies are necessary to adjust the B field profile for highest positron yield.
 In section~\ref{sec:OMDdesign} the possible design of a pulsed matching system   and its influence on the spinning target is discussed.

 \section{Design of an Optical  Matching Device: A pulsed solenoid}\label{sec:OMDdesign} 
A special feature of the undulator driven positron source is the 1\,ms long photon pulse, incident on the rotating wheel. This is much longer than usually used for conventional positron sources with micro-second pulses where Flux concentrators (FC) are convenient for such beam structures. However, due to time varying skin effect in these FC, when driven with ms-pulses, the magnetic field will strongly vary during the beam pulse~\cite{ref:e+WG,ref:GaiLiu-OMD}.
To circumvent this problem for the ILC, a basically simple solenoid , wound in a conical shape (see figure~\ref{fig:B+solenoid}) is considered; see also reference~\cite{ref:sievers-posipol18}.  
Pulsed solenoids as   adiabatic matching device (AMD) have been used in the past, as \eg{} for the positron source of CERN-LEP.  To achieve multi-Tesla fields, currents of about 50 kA or above are required. Clearly, with such high currents, the solenoid cannot be driven in a d.c. mode, but has to be pulsed. Considering a Cu-conductor of about 1\,cm$\times$1\,cm,  the pulse duration has to be chosen long enough, so that at the peak of the half-sine pulse, eddy current have died out and a stable field over the duration of the beam pulse  of 1\,ms is achieved.  
Such a stability will be reached with a half sine current pulse  with a duration of about 4 ms, where the skin depth in Cu will be about 0.6\,cm, sufficiently larger than the average radius of the conductor  with a resistivity of $4\times 10^{-8}\,\Omega$m.  This pulsing will still provide a large reduction of the electrical power consumption of the solenoid with a duty cycle of only 1\%, when driven at 5\,Hz. In table~\ref{tab:sol-design}  some design figures are quoted.

\begin{table}[h]
\begin{tabular}{|l|c|}
\hline
Half sine pulse duration	&4\,ms\\	
Peak current	&50\,kA	\\	
Repetition  rate &	5\,Hz	\\
Average electrical power &6 kW	\\
Water cooling flow &0.17\,l/s	\\
Temperature rise in cooling water&	9\,K	\\
\hline
Peak magnetic field	&5.2\, T\\
Field at target &	3\,T	\\
Field at target with upstream booster coil	&4\,T	\\
Stress due to magnetic field &	$\le 40\,$MPa	\\
\hline
Beam induced effects at entrance of the solenoid, r=1\,cm	&PEDD   
 13\,J/g		\\
Average beam power deposition &	600\,W/cm$^3$\\	
Thermal stress& $\approx 100\,$MPa	\\
displacement per atom (dpa)&	0.15/5000\,h\\
\hline 					
\end{tabular}
\caption{ Design figures for a conical pulsed solenoid, as sketched in figure~\ref{fig:B+solenoid}. The inputs for the beam induced effects were provided by T. Takahashi (Univ. Hiroshima) and A. Ushakov (Hamburg University).}
\label{tab:sol-design}
\end{table}

As can be seen in table~\ref{tab:sol-design}, the pulsing and cooling of the envisaged pulsed solenoid is feasible. However, detailed engineering is required to provide radiation resistant ceramic insulations between the windings as well as solid mechanical clamps, applied to the coil, to retain the pulsed forces, stresses and vibrations.

\begin{figure}[hp]
\center
 \includegraphics*[width=80mm]{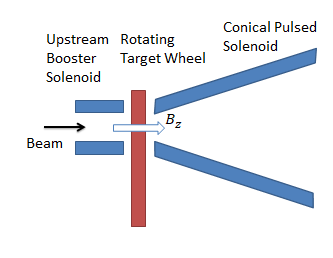}
  \caption{By adding an upstream Booster Solenoid, powered in series with the downstream conical solenoid, an axial field, as indicated by the arrow, of up to 4\,T can be 'dragged' through the 
target.   } 
 \label{fig:booster}
\end{figure}

As shown in table~\ref{tab:sol-design}, for small apertures of 2\,cm diameter at the entrance of the solenoid, the effects induced by the beam at this entrance, become relevant. Thermal fatigue and radiation damage may occur. Therefore, at this stage, larger diameters of the entrance of about 3\,cm or above should also be considered. To compensate for the loss of field, a booster coil placed upstream of the target could recuperate part of the field loss, as indicated in figure~\ref{fig:booster}.

\begin{figure}[hp]
\center
 \includegraphics*[width=120mm]{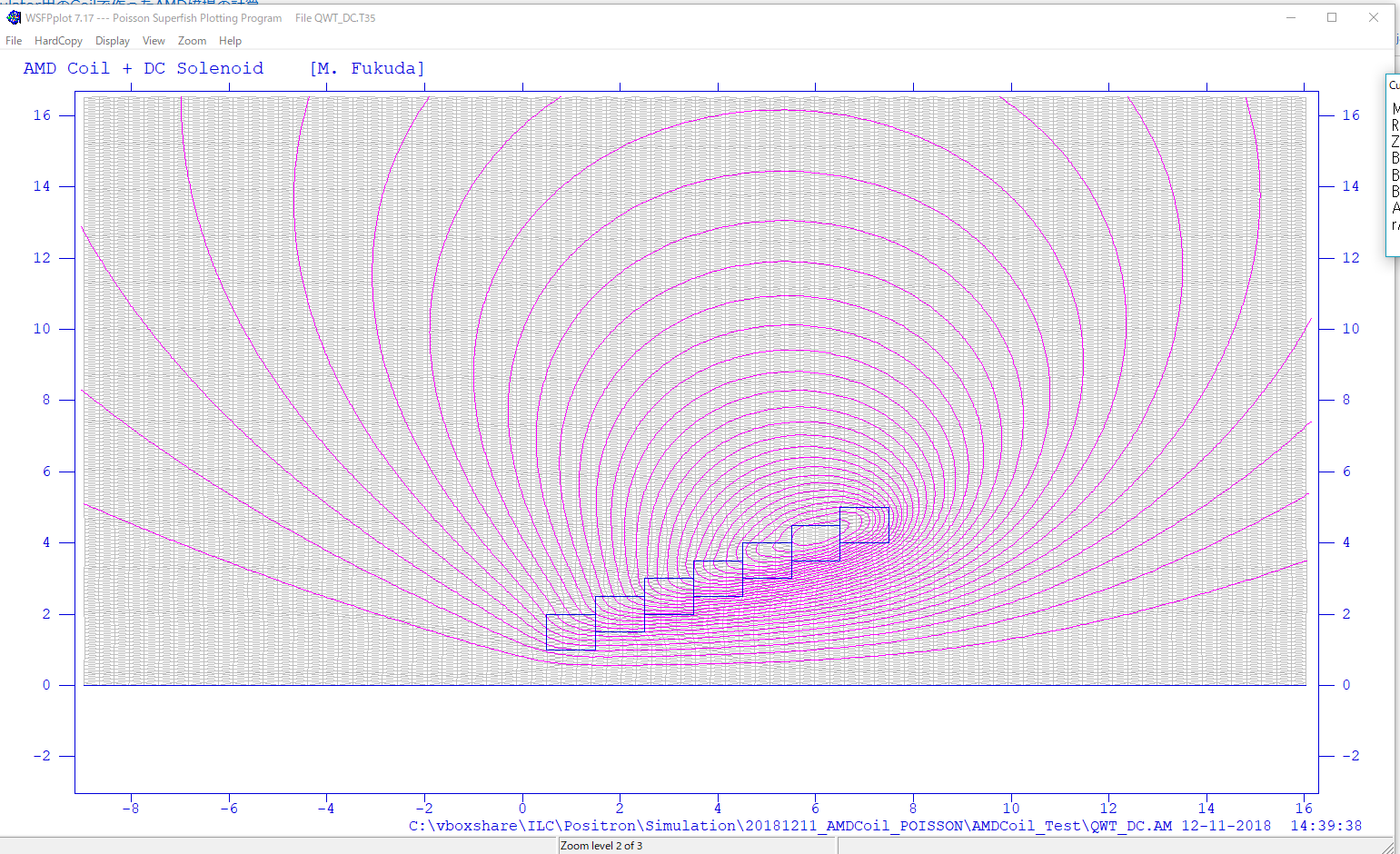}
  \caption{Magnetic field in the conical solenoid used for first  positron yield calculations~\cite{ref:fukuda-solenoid}.  } 
 \label{fig:fukuda-poisson}
\end{figure}

Some very preliminary calculations of the positron yield have been made by M. Fukuda (KEK)~\cite{ref:fukuda-solenoid}  
for the conical solenoid as illustrated in figure~\ref{fig:fukuda-poisson}.  
The increase of the yield is striking. Therefore, further studies of this scheme at even higher magnetic fields should be pursued.

\subsubsection{	Magnetic effects from the pulsed solenoid on the rotating wheel}\label{sec:magneff}
\begin{figure}[h]
\center
 \includegraphics*[width=90mm]{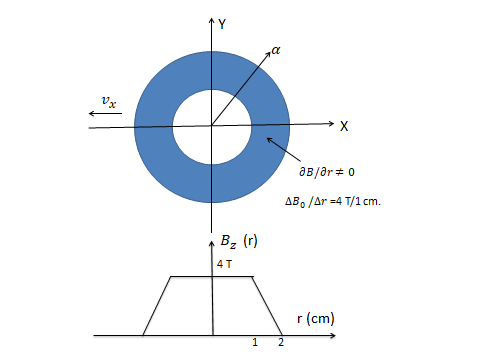}
  \caption{Geometry of the axial magnetic field  $B_z$ in rotational symmetry which penetrates through the rotating wheel. } 
 \label{fig:Bz}
\end{figure}

In a very basic approach to this problem, one can write down the following equations:
\begin{equation}
\frac{\partial B_z (r)}{\partial t}  = \frac{\partial B_z (r)}{\partial r} \cdot  \frac{\partial r}{\partial x} \cdot \frac{\partial x}{\partial t}  =\frac{\partial B_z (r)}{\partial r} \cos (\alpha ) \cdot v \label{eq:Bz}
\end{equation}
For definitions see figure~\ref{fig:Bz} describing a magnetic field $B_z (r)$ with rotational symmetry. 

Equation~\ref{eq:Bz} describes the field variation to which any point in the rotating target is submitted. In particular, field rises occur only in the 'gradient zone' where $\partial B_z / \partial t \ne 0$. This situation  is equivalent to  a non-rotating, stationary target  submitted   to a linear rising field $\propto t/\tau$ in the 'gradient zone' with a rise time $\tau = \Delta r/ v$ .

From this, the time variation of the magnetic flux, the induced voltage $\partial \Phi / \partial t$ can be estimated by integration of $\partial B/\partial t$  over that surface of the Ti-target,  $\partial \Phi /\partial t =\int \partial B /\partial t\cdot df$, which is submitted to the flux change from the solenoid. Finally, by Ohm's law the current density $ j(r,\alpha)=\sigma \partial \Phi /\partial t / \oint ds$  can be evaluated. This  then results in the power $W$ and braking Lorentz forces $F_L$ to which the rotating wheel is submitted:
\begin{eqnarray}
  W&=&\frac{1}{\sigma}\cdot \int j^2 dV \propto \sigma B_0^2  v^2\,  \nonumber \\
  F_L&=~&\int j B dV ~~\propto \sigma B_0^2 v \, .\nonumber
\end{eqnarray}
As has been pointed out in references~\cite{ref:UK-eddy,ref:smythe}, for Ti with its low electrical conductivity, $\sigma = 5.8\times 10^5\, \Omega^{-1}\mathrm{m}^{-1}$,  and with the time duration for any point of the wheel which it takes  to traverse the non-zero-gradient zone --assumed here  of 1\,cm width-- the field will have a rise time of  $10^{-4}$\,s. This, in turn, will lead to a skin depth of about 1.5\,cm in the Ti-target, which is larger than its thickness of 0.7 cm.  Therefore it is plausible that only induced currents from a slowly varying field can be considered while eddy currents can be ignored. This is also confirmed by more detailed studies in reference~\cite{ref:UK-eddy},   
which clearly demonstrates the above quoted dependencies of the average power and torque from the rotation velocity. The absolute values of $W$ and $F_L$ however, depend on the shape of the magnetic field and the volume of the Ti-target, submitted to this field.

For the magnetic field of $B_0= 4\,$T, assumed in this study (see figure~\ref{fig:Bz}), peak current densities $j$ of about $ 1.9\times 10^7\,$A/m$^2$ have been estimated, which in turn lead to at most  10\,kW of power deposited in the Ti-target, when a d.c. magnetic field is assumed. Now, applying the duty cycle of 1\% due to the pulsed operation of the solenoid, the actual average power deposited in the target, will be around 100\,W. This should be well tolerable, in comparison to the beam power of 2\,kW. 
 
Concerning the pulsed braking forces, they are very pessimistically  given by
\begin{equation}
  F=\frac{10 \mathrm{kW}}{v}=100 \mathrm{N/pulse}\nonumber
\end{equation}
 or a torque of 50\,Nm/pulse,
because in reality  all induced currents contribute to the power while only those in radial direction contribute to the braking. The braking energy per pulse, extracted from the wheel over 4\,ms, amounts to 40\,J/pulse. Relating this to the total kinetic energy stored in the rotating wheel of about  0.5\,MJ, this is large compared to the braking energy per pulse. However, these braking pulses at 5\, Hz will accumulate over time and a slow correction of the rotation velocity within 1--2 s will be necessary to maintain the nominal velocity within the required range of $\pm 2 \times 10^{-4}$. 

As has been discussed in references~\cite{Antipov:2007zzf,Antipov:2007zzc}, a parasitic magnetic field will be created by the currents, induced in the Ti-target. This will superpose and deform the original field from the solenoid. Since most of the currents induced in the target, flow into radial direction (see reference~\cite{ref:smythe}), 
they will create a parasitic transversal field in the x-direction of the velocity, inside and close to the target surface. Taking the above quoted current density $j$ of $1.9\times  10^7\,$A/m$^2$, a transverse field of about 0.14\,T must be expected at the location close to the target. This will create a systematic field deformation, a transverse field component, as well as a field drag of the original field.  According to reference~\cite{Antipov:2007zzf,Antipov:2007zzc}, a correction by a simple dipole field may have to be applied. 
Finally thoughts have to be given to the return field around the solenoid. Its penetration into the rotating wheel may still deposit additional power in the target. An iron flux trap around the solenoid may be required.

\section{The R\&D path towards the undulator driven target}\label{sec:RandD}
Further studies of the mechanical design and response of the beam induced thermal loads are required, also to optimise the geometry of the spinning wheel and of the target, to improve further the evacuation of the power by radiation.

In case of high luminosity upgrade the temperature in the target increases and a higher heat evacuation  efficiency is desired.    The   efficiency of cooling  could be improved further by conducting the beam power from the Ti target  into a radiator which is arranged close to the target and has a higher thermal conductivity and which can tolerate higher temperatures. Among other things, Graphite, copper and high temperature  Ni- and Co-alloys may be considered. These investigations should be done by finite element computations and be validated, as proposed earlier, by simple laboratory test setups.

Clearly, the proposed design of the pulsed solenoid has to be validated in terms of life time and field quality and with emphasis on its radiation resistance.

The quality of the magnetic field must be studied in detail, in particular the increasing transverse field components towards larger radii inside the aperture of the solenoid, as illustrated in figure~\ref{fig:fukuda-poisson}.
As stated above, the magnetic effects from the pulsed solenoid on the spinning wheel can efficiently be simulated by computer codes and initially be benchmarked with a stationary wheel submitted to a fast pulsed dipole magnet. However, a complete  prototype will be necessary to validate the final design.

\section{Polarization of positrons}\label{sec:e+pol}
 
There is a continuing discussion whether  the realization of an electron driven positron source is more easy the the construction of the undulator driven positron source.  However, at the ILC the electron driven source delivers only an unpolarized positron beam. Thus,  the decision about the positron source must include also the relevance of a polarized positron beam  for the intended physics measurements.  
\\
Future high-energy \ee{} linear colliders will probe the Standard Model and physics beyond with excellent precision. Besides linear colliders also circular \ee{} collider projects as FCC-ee and CEPC are under discussion~\cite{ref:FCCee}.  FCC-ee and CEPC could operate at energies up to about 350\,GeV centr-of-mass energy and promise measurements with excellent accuracy below the per mil level. In particular at the Z boson resonance they can easily produce a factor 1000 more events than it would be possible with the so-called GigaZ option of the ILC.

Electroweak interactions do not conserve parity, so beam polarization is essential to measure and to disentangle new phenomena beyond the Standard model (SM).  It is expected that such phenomena can be directly obtained at high collision energies but also by deviations from the SM prediction at lower energies as planned for FCC-ee and CEPC. The lessons from the SLC experiment at the the SLAC linear collider showed that  certain parameters are measured with substantially higher precision if both beams, \ie{} electrons and positrons, are polarized.  Although the 4 LEP experiments collected  a more than  30 times  higher number of Z boson events than SLC,  they  measured the effective weak mixing angle with the same precision. 

What is the situation comparing ILC, in particular ILC250 and GigaZ~\cite{ref:GigaZ}, with FCC-ee and CEPC?  
 High degrees of electron beam  polarization are possible; the ILC e$^-$ beam will be at least 80\% polarized.
Since the generation of an intense (polarized) positron beam is a challenge, simultaneously polarized e$^-$ and e$^+$  beams at linear colliders  are under discussion since many years.
Without going in details as  physics processes and their analyses, the benefit of polarized positron beams is given  by the following reasons (see also references~\cite{MoortgatPick:2005cw,Karl:2017xra,Karl:2019hes}):
\begin{itemize}
\item
There are  4 combinations of e$^+$ and e$^-$ helicity states in the collision of high-energy electrons and positrons. 
Only with both beams polarized  each of these  initial state combinations can be explicitly realized in a collider. 
\item
With the 'right' helicity combination of initial states a higher effective luminosity  is achieved: ${\cal L}_{\rm eff}/{\cal L}=1-(1-P_{e^-}P_{e^+})$.
A  higher  number of specific events is achieved in shorter running time. For example,
assuming $P_\mathrm{e^-}=90\%$ and  $P_\mathrm{e^+}=30\%$ the effective luminosity can be almost a factor  1.3 higher than without positron polarization.  
\item
The suppression of background is crucial for precision measurements. With polarized beams the desired initial states can be enhanced or suppressed.  This improves the discrimination and control of background processes.  
\item
Polarized beams provide a high flexibility to evaluate systematic effects. 
It is very difficult to detect and correct time-dependent effects, correlations or a bias in the polarimeter measurement.  If both beams are polarized, such systematic effects can be much better controlled, and their impact on the uncertainty of observables  can be substantially reduced down to negligible values.
\item
In case of deviations from the Standard Model predictions, polarization of both beams
enhances significantly the possibility to confirm the existence of a new phenomenon: High precision,  flexible configuration of initial states and a larger number of independent observables could even allow to unravel underlying physics.
\item
An independent determination of beam polarization and left-right asymmetries is only possible    if 
both beams are polarized. 
\end{itemize}
One should keep in mind that also the zero polarization of an unpolarized positron beam must be confirmed  
to avoid any bias in the physics analyses~\cite{Fujii:2018mli}.
\\
All these arguments suggest that positron polarization is crucial  for precision measurements at ILC250 and inevitable for GigaZ.  
\section{Summary}\label{sec:sum}

The   positron source based on a helical undulator  is the baseline option for the positron generation at the ILC.   It is preferred  to the electron driven scheme for several reasons:  the power absorption in the target  and the source facilities is substantially lower,  less neutrons  are generated and   the target system is less activated.  Further, the sensitivity to changes of the positron damping ring is  lower. And most important for the precision  measurements and the  ILC physics goal   is the delivery of a positron beam which is at least 30\% polarized.   
 
 The undulator based  ILC positron source has been intensively studied in the past and has undergone various modifications and improvements. It has been demonstrated that this source with the  required performance, and in particular with the required positron yield of 1.5\,e$^+$/e$^-$  is technically feasible. Its essential components consist of the Ti-target with a thickness of 7\,mm for ILC250, shaped into a wheel with a diameter of 1\,m and rotating in vacuum at 2000 rpm. The average beam power of about 2 kW  
which is deposited in the target at nominal luminosity, is evacuated by heat radiation from the surface of the wheel into a stationary, water cooled heat sink.

Due to the particular long photon beam pulse of 1 ms at 5 Hz,  for the positron collection system, an efficient Adiabatic Matching Device, AMD, is required. Flux concentrators as used elsewhere,   
\ie{} SLAC, SKEKB,  are operated at micro-s pulse durations; they are not well suited for the long beam pulse of 1\,ms.
In reference~\cite{ref:e+WG},  the positron collection system for the undulator scheme with long beam pulses has been stated as one of the open issues. Here, a possible technical solution is proposed, by which this issue can be resolved by a pulsed solenoid collector.
A solenoid driven with pulses of 4 ms duration and at 5 Hz has been studied. Such a device can provide peak fields of about 5 T in the solenoid while 2-3\,T fields can be tolerated at the rear of the Ti-wheel, without excessive additional heating and braking of the spinning wheel. 
Some possible design modifications of the wheel, to improve further its cooling, and the technical R\&D path towards the development of the essential parts of the undulator driven source have been outlined.

\section*{Acknowledgment}
This work was partially   supported by the German Federal Ministry of Education and Research,
Joint Research Project R\&D Accelerator "Positron Sources'', Contract Number 05H15GURBA.
The fruitful discussions on the AMD with M. Fukuda/KEK, L. Rinolfi/CERN and R. Chehab/LAL are gratefully acknowledged. 
We thank Felix Dietrich for his contributions to the positron target project.

\bibliography{lcws19-SRiemann.bib-old2}

\begin{thebibliography}{10}

\bibitem{TDR1}
{T. Behnke et al.}
\newblock {The International Linear Collider Technical Design Report - Volume
  1: Executive Summary}.
\newblock 2013.

\bibitem{TDR31}
{C. Adolphsen et al.}
\newblock {The International Linear Collider Technical Design Report - Volume
  3.I: Accelerator \& in the Technical Design Phase}.
\newblock 2013.

\bibitem{TDR32}
{C. Adolphsen et al.}
\newblock {The International Linear Collider Technical Design Report - Volume
  3.II: Accelerator Baseline Design}.
\newblock 2013.

\bibitem{ref:LCWS18}
{F. Dietrich, G. Moortgat-Pick, S. Riemann, P. Sievers, A. Ushakov}.
\newblock {Status of the undulator-based ILC positron source}.
\newblock In {\em {International Workshop on Future Linear Colliders (LCWS
  2018) Arlington, Texas, USA, October 22-26, 2018}}, 2019.

\bibitem{ref:e+WG}
{K.~Yokoya (ed.), Positron Working Group W.~Gai et al}.
\newblock {Report on the ILC Positron Source}, May 23 2018.
\newblock {https://edmsdirect.desy.de/item/D00000001165115}.

\bibitem{Omori:2011wq}
{T. Omori et al}.
\newblock {A conventional positron source for International Linear Collider}.
\newblock {\em Nucl. Instrum. Meth.}, A672:52--56, 2012.

\bibitem{Nagoshi:2020blm}
{H. Nagoshi et al}.
\newblock {A design of an electron driven positron source for the
  internationallinear collider}.
\newblock {\em Nucl. Instrum. Meth.}, A953:163134, 2020.

\bibitem{Aihara:2019gcq}
H.~Aihara et~al.
\newblock {The International Linear Collider. A Global Project}.
\newblock 2019.

\bibitem{ref:undproto}
D.~J. et~al Scott.
\newblock {Demonstration of a High-Field Short-Period Superconducting Helical
  Undulator Suitable for Future TeV-Scale Linear Collider Positron Sources}.
\newblock {\em Phys. Rev. Lett.}, 107:174803, Oct 2011.

\bibitem{ref:LCWS17proc}
{S. Riemann, F. Dietrich, G. Moortgat-Pick, P. Sievers, A. Ushakov}.
\newblock {The ILC positron target cooled by thermal radiation}.
\newblock In {\em {International Workshop on Future Linear Collider (LCWS2017)
  Strasbourg, France, October 23-27, 2017}}, 2018.

\bibitem{ref:posipol16-AU}
A~Ushakov.
\newblock
  {\href{https://indico.lal.in2p3.fr/event/3288/contributions/8120/attachments/7640/9004/Ushakov-POSIPOL2016.pdf}{Talk}
  given at POSIPOL, August 2016, Orsay, France}.

\bibitem{ref:AU-250GeVthickness}
{A. Ushakov, V. Kovalenko, G. Moortgat-Pick, S. Riemann, F. Staufenbiel}.
\newblock {Simulations of the ILC Positron Source at Low Energies}.
\newblock In {\em {Proceedings, 4th International Particle Accelerator
  Conference (IPAC 2013): Shanghai, China, May 12-17, 2013}}, page TUPME003,
  2013.

\bibitem{ref:AU-OMDyield}
{A. Ushakov}.
\newblock {Positron Yield Calculations for the Undulator Based Source at 250
  GeV CM Energy}.
\newblock
  {\href{https://agenda.linearcollider.org/event/7826/contributions/41436/attachments/33054/50352/Ushakov-ALCW18.pdf}{talk}
  given at the Asian Linear Collider Workshop ALCW2018 in Fukuoka, Japan, 28th
  May -- June 2nd, 2018}.

\bibitem{ref:Gronberg-FC}
{J. Gronberg et al}.
\newblock {Prototyping of the ILC Baseline Positron Target}.
\newblock In {\em {International Workshop on Future Linear Colliders (LCWS11)
  Granada, Spain, September 26-30, 2011}}, 2012.

\bibitem{ref:AU-OMD}
{A. Ushakov, G. Moortgat-Pick, S. Riemann}.
\newblock {Undulator-Based Positron Source at 250 GeV CM Energy with Different
  Optical Matching Devices: Pulsed Flux Concentrator and Quarter Wave
  Transformer}.
\newblock In {\em {International Workshop on Future Linear Collider (LCWS2017)
  Strasbourg, France, October 23-27, 2017}}, 2018.

\bibitem{ref:GaiLiu-OMD}
{W. Liu, W. Gai, L. Rinolfi and J. Sheppard}.
\newblock {An Undulator based Polarized Positron Source for CLIC}.
\newblock {\em Conf. Proc.}, C100523:THPEC035, 2010.

\bibitem{ref:Gronberg-posipol13}
{J. Gronberg et al}.
\newblock
  \href{https://indico.fnal.gov/event/ANLHEP114/session/8/contribution/25/material/0/0.pdf}{Talk}
  given at POSIPOL Workshop, 4-6 September 2013, Argonne National Laboratory,
  USA.

\bibitem{ref:sievers-posipol14}
P.~Sievers.
\newblock
  \href{https://kds.kek.jp/indico/event/15241/session/17/contribution/12/material/slides/0.ppt}{Talk}
  given at POSIPOL, August 2014, Ichinoseki, Japan.

\bibitem{ref:sievers-posipol16}
P.~Sievers.
\newblock
  \href{https://indico.lal.in2p3.fr/event/3288/contributions/8167/attachments/7656/9024/POSIPOL-2016_Peter.pptx}{Talk}
  given at POSIPOL Workshop, 14-16 September 2016, LAL Orsay, France.

\bibitem{Mills:thermo}
{K.C. Mills}.
\newblock {Recommended Values of Thermophysical Properties for Selected
  Commercial Alloys}.
\newblock 1st edition, 2002, eBook ISBN: 9781845690144, Woodhead Publishing.

\bibitem{ref:ANSYS}
ANSYS.
\newblock \href{http://www.ansys.com}{http://www.ansys.com}.

\bibitem{Ushakov:2017dha}
{A. Ushakov et al}.
\newblock {Material Tests for the ILC Positron Source}.
\newblock In {\em {Proceedings, 8th International Particle Accelerator
  Conference (IPAC 2017): Copenhagen, Denmark, May 14-19, 2017}}, page
  TUPAB002, 2017.

\bibitem{Heil:2017ump}
{P. Heil, K. Aulenbacher, T. Beiser et al}.
\newblock {High Energy Density Irradiation With MAMI LINAC}.
\newblock In {\em {Proceedings, 8th International Particle Accelerator
  Conference (IPAC 2017): Copenhagen, Denmark, May 14-19, 2017}}, page
  TUPAB003, 2017.

\bibitem{ref:AU-posipol18}
{A. Ushakov}.
\newblock {Acceptable Peak Temperature and Thermal Stress in Ti6Al4V Target of
  ILC Positron Source}.
\newblock
  \href{https://indico.cern.ch/event/727621/contributions/3109101/attachments/1709300/2755120/Ushakov-POSIPOL18.pdf}{Talk}
  given at POSIPOL Workshop, 3-5 September 2018, Geneva, Switzerland.

\bibitem{ref:sievers-posipol18}
P.~Sievers.
\newblock
  \href{https://indico.cern.ch/event/727621/contributions/3109115/attachments/1708258/2754534/POSIPOL_2018-2.pdf}{Talk}
  given at the POSIPOL Workshop, 3-5 September 2018, Geneva, Switzerland.

\bibitem{ref:fukuda-solenoid}
{M. Fukuda}.
\newblock {private communication, 2019}.

\bibitem{ref:UK-eddy}
{I. Bailey et al}.
\newblock {Eddy Current Studies From the Undulator-based Positron Source Target
  Wheel Prototype}.
\newblock {\em Conf. Proc.}, C100523:THPEC033, 2010.

\bibitem{ref:smythe}
W.R. Smythe.
\newblock {On eddy currents in a rotating disk}.
\newblock {\em Electrical Engineering}, 61 no. 9:681--684, Sept. 1942.

\bibitem{Antipov:2007zzf}
{S. Antipov, L. Spentzouris, W. Liu, W. Gai}.
\newblock {Numerical studies of International Linear Collider positron target
  and optical matching device field effects on beam}.
\newblock {\em J. Appl. Phys.}, 102:014910, 2007.

\bibitem{Antipov:2007zzc}
{S. Antipov, L. K. Spentzouris, W. Liu, W. Gai}.
\newblock {Simulations of the Rotating Positron Target in the Presence of OMD
  Field}.
\newblock {\em Conf. Proc.}, C070625:2909, 2007.
\newblock [2909(2007)].

\bibitem{ref:FCCee}
A.~Abada et~al.
\newblock {FCC-ee: The Lepton Collider}.
\newblock {\em Eur. Phys. J. ST}, 228(2):261--623, 2019.

\bibitem{ref:GigaZ}
Kaoru Yokoya, Kiyoshi Kubo, and Toshiyuki Okugi.
\newblock {Operation of ILC250 at the Z-pole}.
\newblock 2019.

\bibitem{MoortgatPick:2005cw}
{G. Moortgat-Pick et al}.
\newblock {The Role of polarized positrons and electrons in revealing
  fundamental interactions at the linear collider}.
\newblock {\em Phys. Rept.}, 460:131--243, 2008.

\bibitem{Karl:2017xra}
{R. Karl and J. List}.
\newblock {Polarimetry at the ILC}.
\newblock In {\em {Proceedings, International Workshop on Future Linear
  Colliders 2016 (LCWS2016): Morioka, Iwate, Japan, December 05-09, 2016}},
  2017.

\bibitem{Karl:2019hes}
{R. Karl}.
\newblock {\em {From the Machine-Detector Interface to Electroweak Precision
  Measurements at the ILC -- Beam-Gas Background, Beam Polarization and Triple
  Gauge Couplings}}.
\newblock PhD thesis, Hamburg U., Hamburg, 2019.

\bibitem{Fujii:2018mli}
{K. Fujii et al}.
\newblock {The role of positron polarization for the inital $250$ GeV stage of
  the International Linear Collider}.
\newblock 2018.

\end{thebibliography}
\bibliographystyle{unsrt}
\end{document}